# DETECT LANGUAGE OF TRANSLITERATED TEXTS

Sourav Sen[1]

[1]Department of Physics, Duke University, Durham, NC, USA
sourav.sen@duke.edu

## ABSTRACT

*Informal transliteration from other languages to English is prevalent in social media threads, instant messaging, and discussion forums. Without identifying the language of such transliterated text, users who do not speak that language cannot understand its content using translation tools. We propose a Language Identification (LID) system, with an approach for feature extraction, which can detect the language of transliterated texts reasonably well even with limited training data and computational resources. We tokenize the words into phonetic syllables and use a simple Long Short-term Memory (LSTM) network architecture to detect the language of transliterated texts. With intensive experiments, we show that the tokenization of transliterated words as phonetic syllables effectively represents their causal sound patterns. Phonetic syllable tokenization, therefore, makes it easier for even simpler model architectures to learn the characteristic patterns to identify any language.*

## 1. INTRODUCTION

In the age of instant messaging and social media, one often can find texts transliterated from other languages in English. Computer and mobile phone keyboards being mostly in English might make it convenient for non-english speakers to use latin scripts for communicating in their language, instead of typing in their original scripts, especially in informal settings like social media threads, chats and discussion forums. There may exist formal systems for some languages, see [1], [2], [3], to convert text from their original script to latin script, referred to as 'romanization'. However, in informal settings, people often do not follow such formal systems. Such transliterated texts become encrypted, in a way, to others who don't speak or recognize that language.

In this paper, we approach the problem of identifying the language of informal transliterations in English. For this work, we designed a classifier to identify transliterations in Bengali and Korean languages. We conjecture that every language has its characteristic sound patterns, which can be used to identify them. We propose that tokenizing the transliterated words into phonetic syllables help LID models extract these characteristic sound patterns easily and identify the language of these transliterated words. Our work can be summarized as follows: (1) We design a single layer unidirectional LSTM recurrent neural network and train on the phonetic syllable tokenizations of transliterated words. (2) We extensively investigate the efficacy of the phonetic syllable tokenization of transliterated words in representing the underlying sound patterns.

The paper is structured in the following way: In section 2, we introduce related work in detecting languages from texts. We present our approach to the solution in section 3 and evaluate it on extensive experiments in section 4. We conclude our work in section 5.

## 2. RELATED WORK

There are very few tools online which can identify the language of transliterations correctly and that too for a handful of languages only, as in [4]. Perhaps, it is the scarcity of parallel corpus for transliterated and original script texts in other languages that restricts the inclusion of those languages in the LID tools for transliterations.

The state-of-the-art Language identification systems for texts use LSTM based recurrent NNs to identify the language. The studies in [5] have shown bi-directional LSTMs to perform reasonably well for identifying the languages of short strings, where each language shares a common script. Recognizing the correct script narrows down the probable languages of the text. It is, however, not possible for transliterated texts where the script is different from its original one. Therefore, it would require a lot more data to train a language identification model for transliterated texts, making it hard to develop such systems for less commonly transliterated languages.

Owing to these complications, we tried to delve deeper to understand how one recognizes languages without the help of scripts. A closer realm of such an objective is language identification in speech. We hypothesize that with enough exposure to any language, one is able to identify the language, even without understanding it, since every language has some characteristic sound patterns that our brains are able to learn. Since pattern recognition is a specialization of machine learning models, automatic speech recognition (ASR) systems have been using recurrent neural network-based models to identify languages from audio inputs.

The ASR systems extract characteristic sound patterns from the audio clips using spectrograms (see [6]). Traditional ASR LID systems, in [7], [8], have been using LSTM based models with hand-crafted MFFC-SDC features from the spectrograms. Bartz et al. in [9] have developed an end-to-end LID system by extracting the input features from the spectrograms using CNNs (and Google's Inceptron-v3, see [10]) and passed them to LSTM based architecture to identify languages, beating many state-of-the-art results.

Since informal transliterations are often done phonetically even if the original language is not strictly phonetic, analogous features of transliterated texts to the spectrograms in audio files would be phonetic syllables. We used this observation as a basis to design our solution to this problem, described in the Proposed Solution section.

## 3. PROPOSED SOLUTION

We tokenize the transliterated words into phonetic syllables using an open-source hyphenation package called 'Pyphen' (see [11]). This package uses hyphenation libraries to split words into syllables. Since English is a pseudo-phonetic language, we used the hyphenation scheme of the Italian language, which is strictly phonetic and also uses Latin scripts, to split the words. The phonetic syllable tokenized words are then inputted to an ML model.

### 3.1. Data Collection

We used Korean and Bengali song lyrics online transliterated in English as our training corpus. Figure 1 shows that Bangla words generally consist of two phonetic syllables. Figure 2 shows that the total length of phonetic syllables in both languages is less than ten characters.

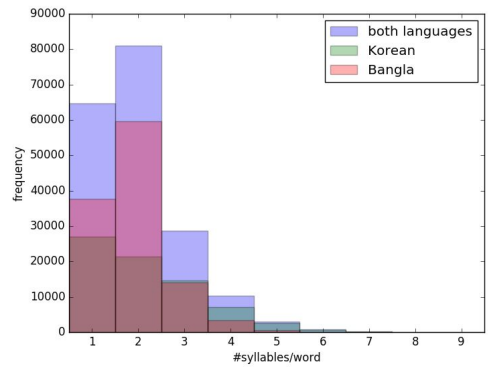

Figure 1. Distribution of phonetic syllables per word in Bangla and Korean transliterations.

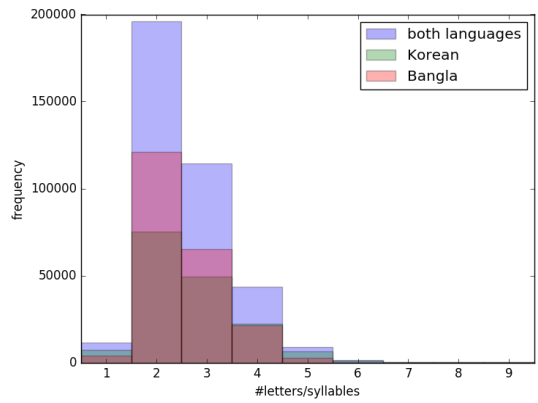

Figure 2. Distribution of length (number of characters) of phonetic syllables in Bangla and Korean transliterations.

### 3.2. Architecture

We used a unidirectional single-layered LSTM architecture for our LID ML model. Each token, i.e., phonetic syllable of a word, is converted into an integer using the MD5 hashing function. The sequence of integers is then inputted by the ML model. The first layer of the ML model is an embedding layer which outputs 8-dimensional word embedding vectors. This layer feeds its output to the LSTM unit, which, at the end of the input sequence, outputs the hidden state to a fully-connected layer with one output unit serving as the classifier. Figure 3 provides a schematic overview of the network architecture.

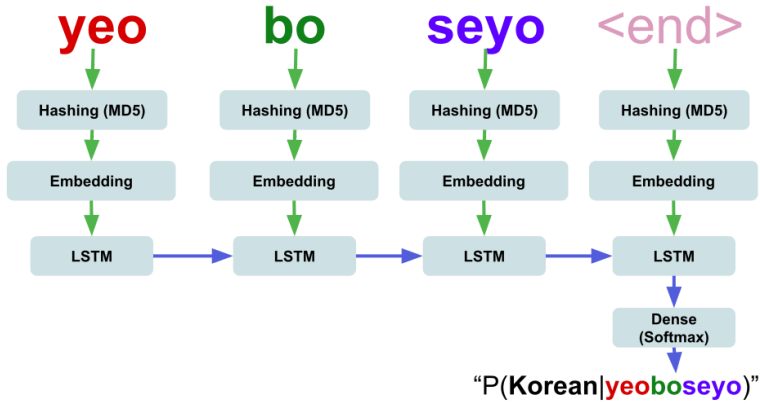

Figure 3. The architecture of our proposed Language Identification System for transliterated systems, using phonetic syllables as input. The phonetic syllables are converted to integers using MD5 hashing function followed by an embedding layer and fed to LSTM at every time step. The final LSTM output is fed into a fully-connected layer for classification.

### 3.3. Baseline Approach

To benchmark the performance of our approach, we used characters as tokens for the transliterated words. The characters were then alphabetically mapped to integers. The sequence of integers is then fed to the same architecture described in section 3.1.

## 4. EXPERIMENTS

Using the dataset and the network architecture introduced in section 3, we conducted several experiments to assess the performance of our proposed solution. While performing our experiments, we had a range of different questions in mind:
- Does our solution of tokenizing the words into phonetic syllables give better performance accuracy when compared to the baseline approach where the words are tokenized using their characters?
- Is the network able to reliably discriminate between languages?
- Are phonetic syllables robust representations of the underlying sound patterns when subjected to noise?

### 4.1. Environment

We implemented our proposed model using Keras (see [12]) with the Tensorflow [13] backend. The training and test words were randomly sampled from each corpus, with their respective sizes given in Table 1. We split the training data into training (90% ), and a validation (10%) set.

Table 1. Data split between training (and validation) and test set.

| Data Set | Bangla | Korean |
|---|---|---|
| Training (& validation) data | 50,000 words | 50,000 words |
| Test data | 10,000 words | 10,000 words |

For training our networks, we used the Adam (see [14]) optimizer. We used regularization techniques such as L2 regularization, early stopping, and drop out to prevent any overfitting. We observed the following metrics: accuracy, the area under the ROC curve (AUC). We refer to the baseline approach in section 3.3 as the baseline in this section.

## 4.2. Performance with Phonetic Syllables

Our solution approach of tokenizing the transliterated words using phonetic syllables gave an AUC of 0.984 (see Figure 4) with an accuracy of 94% (decision boundary at 0.5), which was about 50% higher in performance than the baseline approach (see Table 2).

Table 2. Performance comparison between our proposed LID (using phonetic syllable) with the baseline LID (using individual letters or characters).

| Model | Accuracy |
|---|---|
| Phonetic syllables | 0.94 |
| Letters (baseline) | 0.52 |

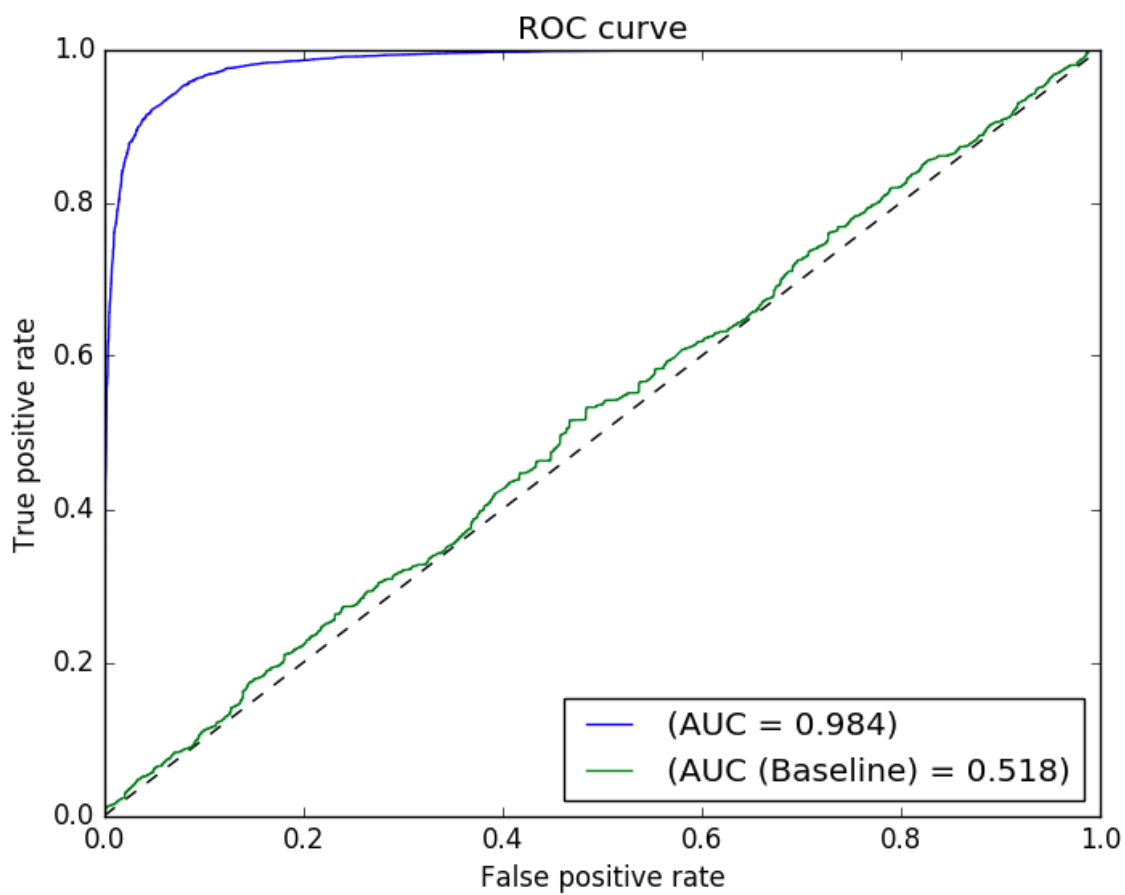

Figure 4. ROC curve of our proposed LID using phonetic syllables (in blue) and the baseline LID using characters (in green).

### 4.3. Noise Robustness

The significantly better performance with phonetic syllables in section 4.2 indicates that phonetic syllables are, in some ways, able to represent the underlying sound patterns corresponding to the pronunciation of the words. We further test the efficacy of these phonetic syllables in representing sound patterns, used to pronounce them, by adding noise or variations in the spellings of the test words. An example of such variation can be seen in Figure 5.

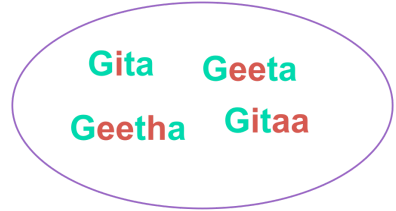

Figure 5. Ambiguity in the phonetical spelling of transliterated Sanskrit word *gītā*.

One subject to such variations, the prediction scores of our trained model (with phonetic syllables), should only perturb a little if the phonetic syllables are able to represent the sound patterns to the ML model efficiently. The effect of such variations on the performance of our LID is quite directly relevant, as such variations often happen in informal transliterations with no fixed convention for spelling words.

#### 4.3.1. Simulation

We used a toy simulation to introduce variations in the spelling of the test words. Since most common places where people vary the spelling of transliterated words are in the usage of contiguous vowels (also seen in the example in Figure 5), we vary the count of vowels. For this variation, we scan the words for vowels. Once a vowel is found, we draw an integer *count* from a random number generator with discrete uniform distribution U(0, N) and replace that vowel in the word by a string of *count* number of copies of the vowel. Here, N is a parameter, which can be used to control the extent of perturbation.

#### 4.3.2. Metrics For Measuring Perturbation Effects

We studied the change in the correct class score due to perturbation. The correct class score is the score predicted by our LID for the correct language of the word, for examples, if a Bangla test word *b* got a score P(*b*|Bangla) by our LID as 0.2, and a Korean word *k* got a score P(*k*|Korean) as 0.2, both words *b* and *k* have their correct class scores as 0.2, respectively. The correct class score before and after the perturbation, when N=3, is shown in Figure 6.

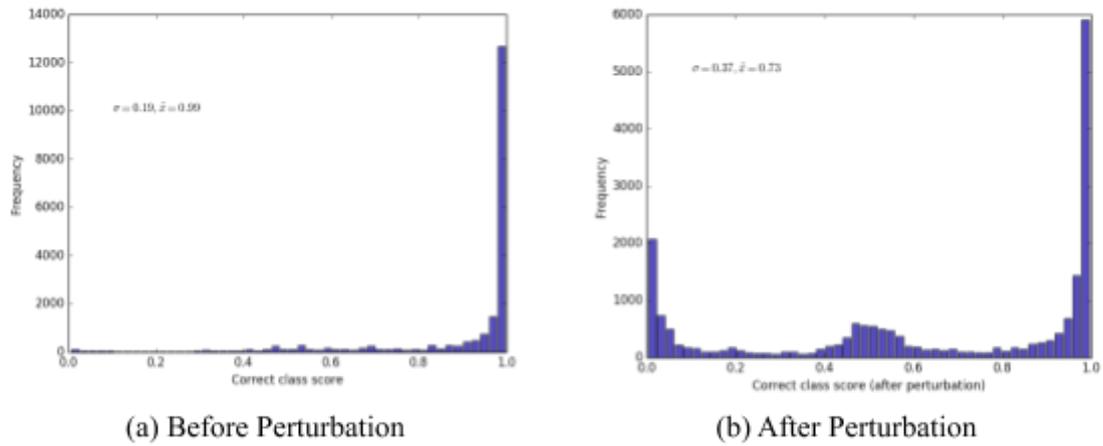

(a) Before Perturbation  (b) After Perturbation

Figure 6. The correct class score distribution of test words before and after variation in spelling. Here perturbation parameter N=3.

We use the standard deviation of the fractional difference ($\sigma_{correctscore}$) in the correct class score distribution (shown in Figure 7) to quantify the variation in the correct class score. One can interpret the $\sigma_{correctscore}$ as the average resolution for a word before and after perturbation.

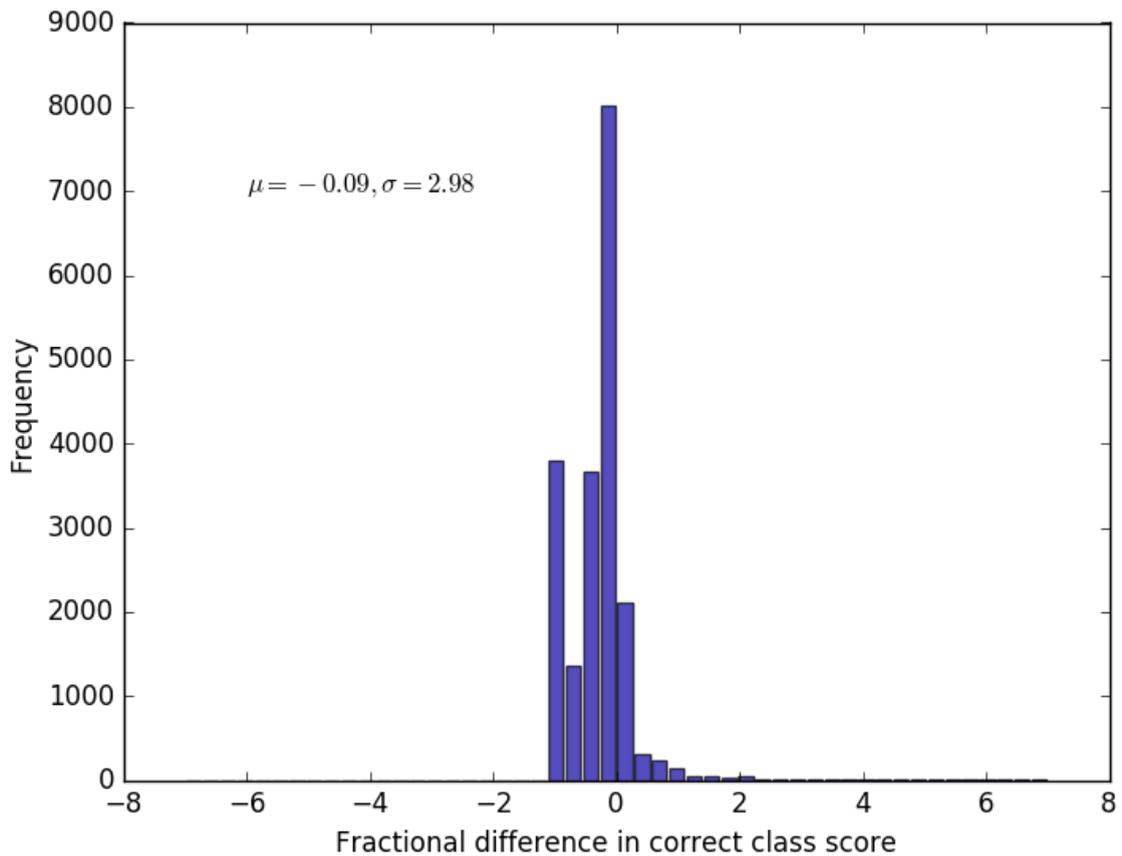

Figure 7. Distribution of the fractional difference in the correct class score ( $= \frac{(\text{perturbed correct class score} - \text{original correct class score})}{\text{original correct class score}}$ ) due to variation in spelling of the test words. Here perturbation parameter N=3.

We use the modified coefficient of variation $CV_{correctscore}$ (= median/std.dev) of the correct class score distribution before perturbation, shown in Figure 6 (a), to quantify the average resolution between two dissimilar words. If $\sigma_{correctscore}$ is smaller than $CV_{correctscore}$ (or $\sigma_{correctscore}/CV_{correctscore} < 1$) then the words after small perturbation remain similar to the LID than two random words. We, therefore, use $\sigma_{correctscore}/CV_{correctscore}$ as a metric for quantifying the effect of perturbation. A higher $\sigma_{correctscore}/CV_{correctscore}$ value refers to a larger dissimilarity between the perturbed and original version of the words for the LID classifier.

Since the above metric assumes a parametric form of the correct class score and fraction difference in the correct class score distributions, we used a non-parametric approach to define a metric as well. For the non-parametric approach, we use the minimum Mann-Whitney U [refer Mann-Whitney U] score (min. U) between the before (Figure 6 (a)) and after (Figure 6 (b)) correct class score distribution as a metric to measure the effect of perturbation. The Mann-Whitney U score of 0.5 (upper limit) would mean the correct class score distribution of the test words before and after distribution is indistinguishable. While a min. U score of 0 (lower limit) would mean that the correct class score distributions before and after perturbation have no overlap at all, so they are completely dissimilar words to the LID classifier. Both metrics have been evaluated for N=3 in Table 3.

Table 3. Metrics for perturbation effects on the LID for perturbation parameter N=3.

| Metric | Value |
|---|---|
| Coefficient of variation ($CV_{correctscore}$) | 5.15 |
| Fractional variation std.dev ($\sigma_{correctscore}$) | 2.99 |
| $\sigma_{correctscore}/CV_{correctscor}$ | 0.58 |
| min. U | 0.25 |

### 4.3.3. Effect Of Noise On LID Performance

We control the degree of perturbation using parameter N in the simulation (as discussed in section 4.3.1) and measure its effect on the LID using the two metrics defined in section 4.3.3. As shown in Figure 8, the $\sigma_{correctscore}/CV_{correctscore}$ increases while the min. U decreases with the increase in perturbation, i.e., increase in N. This trend suggests that under smaller perturbation in word spellings when the perturbed version is phonetically similar to the original version, the proposed LID system interprets the perturbed word as similar to the original word. This indicates that the phonetic syllables are able to represent the underlying sound patterns to the sounds to the LID, making them robust to small variations in spellings that occur during informal transliterations.

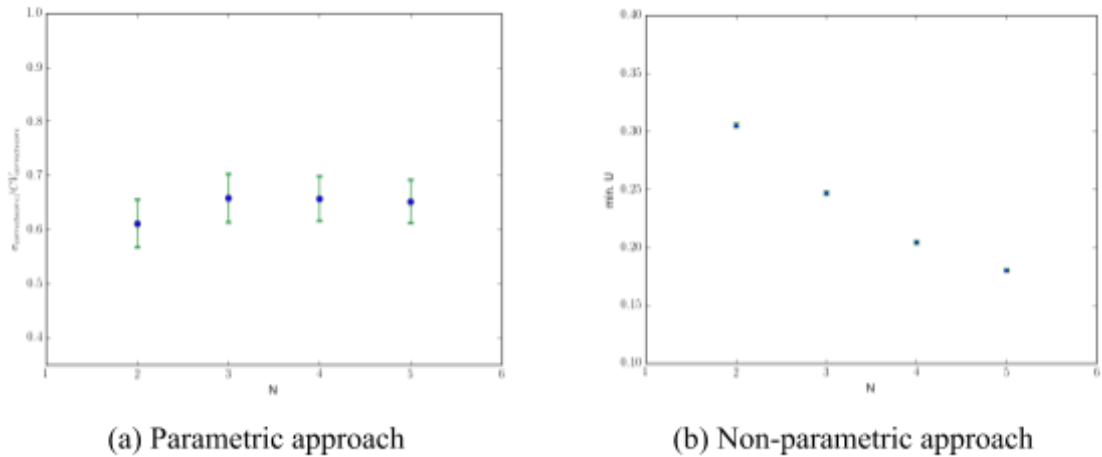

(a) Parametric approach      (b) Non-parametric approach

Figure 8. Effect of perturbation in test word spelling on the performance of our LID system. The $\frac{\sigma_{correctscore}}{CV_{correctscore}}$ metric in (a), defined parametrically, increases while the min.U metric in (b), defined non-parametrically, decreases on increasing the perturbation, i.e., N. Both the trends consistently indicate that under small variation in spelling, phonetically similar words are interpreted as similar words by the LID system as the LID assigns them similar scores.

## 5. CONCLUSION

In this paper, we trained a LID system with phonetic syllables as input features for informal Bangla and Korean transliterations. We used a single-layered LSTM classifier architecture, which performs significantly better with phonetic syllables (~94% prediction accuracy) than with individual characters as inputs. We simulated the spelling variations in test examples to understand the effectiveness of phonetic syllables in characteristic sound pattern representations. We used parametric and non-parametric approaches to define metrics for quantifying the classifier's response to spelling variations. Both metrics indicate that the classifier has some sensitivity to the similarity of spellings within small perturbations. This is a meaningful step forward, and further investigations can reveal if phonetic syllables can help build universal LID systems for informal transliterations (with transfer learning approaches).

**Authors**

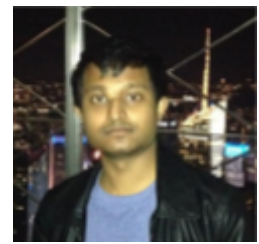

I'm Sourav Sen. I am a Ph.D. student in experimental Particle Physics at Duke University. I work in the ATLAS experiment of the Large Hadron Collider (LHC). My research involves analyzing data from sub-atomic particle collisions at very high energies. I am also very interested in machine learning. In my spare time, I like reading about ML algorithms and also applying them, even beyond my physics research, in some of my other projects.